# Magnetism of $Al_xFe_{2-x}GeO_5$ with Andalusite Structure

K. Kakimoto[1*], S. Takada[1], H. Ohta[2], Y. Haraguchi[1],
M. Hagihala[3‡], S. Torii[3], T. Kamiyama[3*],
H. Mitamura[4], M. Tokunaga[4],
A. Hatakeyama[1], and H. Aruga Katori[1,5†]

[1]*Graduate School of Engineering, Tokyo University of Agriculture and Technology,
Koganei, Tokyo 184-8588, Japan*
[2]*Department of Molecular Chemistry and Biochemistry,
Faculty of Science and Engineering, Doshisha University,
Kyotanabe, Kyoto 610-0321, Japan*
[3]*Neutron Science Division, Institute of Materials Structure Science,
High Energy Accelerator Research Organization,
Tokai, Ibaraki 319-0016, Japan*
[4]*International MegaGauss Science Laboratory,
Institute for Solid State Physics, University of Tokyo,
Kashiwa, Chiba 277-8581, Japan*
[5]*Research Center for Thermal and Entropic Science,
Graduate School of Science, Osaka University,
Toyonaka, Osaka 560-0043, Japan*

The magnetism of $Al_xFe_{2-x}GeO_5$ from $x = 0.09$ to $x = 0.91$ with andalusite structure was examined. The magnetic properties of $Al_xFe_{2-x}GeO_5$ at low temperatures were found to be weak ferromagnetic-like state for $x < 0.3$ and spin-glass state for $x > 0.3$. The small spontaneous magnetization that appears in the weak ferromagnetic-like phase would be due to the presence of Dzyaloshinsky–Moriya interaction or to the difference in the magnitude of the magnetic moment of $Fe^{3+}$ in the octahedral and trigonal bipyramidal sites. The appearance of the spin-glass phase indicates that the dilution of Fe ions by Al ions in $Al_xFe_{2-x}GeO_5$ causes the competition between ferromagnetic and antiferromagnetic interactions. The $x$ dependence of the site occupancy of Fe ions suggests that $Fe_2GeO_5$ with andalusite structure cannot be synthesized.

## 1. Introduction

A polymorph is a material having the same chemical composition but multiple different crystal structures. The stability of the crystal structure depends on the conditions of temperature and pressure during synthesis. $Al_2SiO_5$ comprises three polymorphs whose structures are kyanite, andalusite, and sillimanite.[1-3] Each polymorph crystallizes depending on temperature and pressure during formation. In the synthesis temperature and pressure diagram, kyanite-, andalusite-, and sillimanite-phase are recognized to appear at low temperature and high pressure, high temperature and low pressure, and high temperature and high-pressure phases, respectively. Figure 1 shows the crystal structure of andalusite-, kyanite-, and sillimanite-type $Al_2SiO_5$. Andalusite-type $Al_2SiO_5$ is crystallized in an orthorhombic structure having the space group of *Pnnm* as shown in Fig. 1(a). There are two apparently different Al sites: Al1 is surrounded by six oxygens to form the $AlO_6$ octahedron, and Al2 is surrounded by five oxygens to form the $AlO_5$ trigonal bipyramid. The $AlO_6$ octahedra share their edges to form a one-dimensional chain toward the *c*-axis direction. Neighboring chains are then connected via the vertex-sharing $AlO_5$ trigonal bipyramid. However, as shown in Fig. 1(b), the kyanite-type $Al_2SiO_5$ is crystallized in a triclinic structure having a space group of $P\bar{1}$. There are four types of Al sites: Al1, Al2, Al3, and Al4. Each Al ion is surrounded by six oxygens to form the $AlO_6$ octahedron. The $AlO_6$ octahedra share their edge to form a two-dimensional layer in a *bc* plane. As shown in Fig. 1(c), the sillimanite-type $Al_2SiO_5$ has an orthorhombic structure similar to andalusite-type; its space group is *Pnma*. However, the $AlO_6$ chains along the *c*-axis are connected via the $AlO_4$ tetrahedra.

$Al_2SiO_5$ with either crystal structure can only be synthesized under very high-pressure conditions, while $Al_2GeO_5$, in which Si is replaced by Ge, can be easily synthesized under ambient pressure.[5] In addition, there are $T_2GeO_5$ ($T$ = V, Cr, Fe), materials in which Al ion is

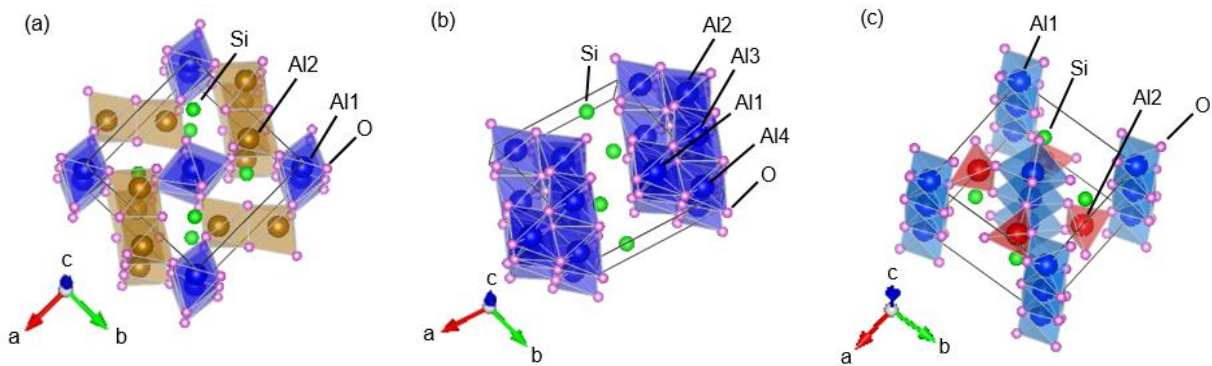

Fig.1 (Color online) Crystal structure of $Al_2SiO_5$ with (a) andalusite, (b) kyanite, and (c) sillimanite structures. The solid black lines show the unit cell of each structure. The VESTA program is used for visualization.[4]

replaced by other trivalent transition metal ions in $Al_2GeO_5$. Among them, $T_2GeO_5$ with $T$ = V,

Cr, Fe are magnetic materials.[6-8] However, despite having the same chemical formula as $Al_2SiO_5$, the synthesis of materials only with kyanite structure has been reported for any magnetic materials $T_2GeO_5$. In contrast, $Al_xFe_{2-x}GeO_5$, in which some of $Al^{3+}$ in $Al_2GeO_5$ is replaced by $Fe^{3+}$, has been reported to have the same polymorphism as $Al_2SiO_5$.[9-12] Al-rich $Al_xFe_{2-x}GeO_5$ crystallizes in either kyanite- or mullite-structure, which is a slightly distorted sillimanite structure, while Fe-rich $Al_xFe_{2-x}GeO_5$ also crystallizes in the andalusite structure.[9] In other words, $Al_xFe_{2-x}GeO_5$ is the only magnetic material with andalusite structure. However, its magnetism has not been investigated in detail; only the samples with $x = 0.58$ and $x = 1.08$ have been reported to be spin-glasses based on their temperature variations of magnetizations.[11,12]

If only the magnetic ions are extracted, as shown in Fig. 2(a), the andalusite structure can be regarded as a structure with identical corrugated lattices aligned parallel to the (110) and ($1\bar{1}0$) planes. As shown in Fig. 2(b), this lattice comprises hexagons and triangles called a bow-tie lattice. Since bow-tie lattices have been the subject of theoretical studies on percolation, it is interesting to investigate the magnetism of $Al_xFe_{2-x}GeO_5$ with andalusite structure from the viewpoint of percolation.

In these circumstances, we report the magnetism of $Al_xFe_{2-x}GeO_5$, the only magnetic material with andalusite structure in the $T_2GeO_5$ series. We also discuss the percolation limit in the andalusite structure compared to that in the bow-tie lattice. Furthermore, from the viewpoint of crystal structure, we investigate whether $Fe_2GeO_5$ corresponding to $x = 0$ in $Al_xFe_{2-x}GeO_5$ can have andalusite structure, i.e., whether $Fe_2GeO_5$ can be polymorphic or not.

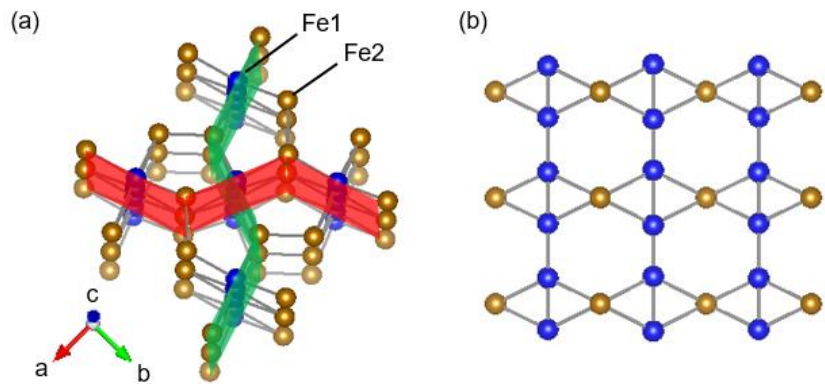

Fig. 2. (Color online) (a) Crystal structure of andalusite, showing only two types of magnetic sites, Fe1 (blue spheres) and Fe2 (brown spheres). The lattices parallel to the (110) and ($1\bar{1}0$) planes are shown in red and green, respectively. (b) bow-tie lattice consisting of Fe1 and Fe2 viewed perpendicular to (110) and ($1\bar{1}0$) planes.

## 2. Experimental

We attempted to synthesize polycrystalline $Al_xFe_{2-x}GeO_5$ with $x \leq 0.91$ using a conventional solid-state reaction method. However, we were unsuccessful in synthesizing the samples with $x < 0.09$ using this method; here, we report on the samples with $0.09 \leq x \leq 0.91$.

Samples were obtained by heating a pelletized mixture of α-$Al_2O_3$, α-$Fe_2O_3$, and $GeO_2$ in a silica tube in an evacuated condition at a certain temperature for 12 h. For the samples with $x = 0.09$ and 0.15, excess $GeO_2$ was added to the mixture considering the evaporation loss. This process prevents the magnetic impurity α-$Fe_2O_3$ from becoming excessive and remaining in the synthesized samples.

The samples were then characterized using powder X-ray diffraction (XRD) on a diffractometer using CuKα radiation. Both lattice parameters and crystal structures were refined as per the Rietveld method using Z-Rietveld.[13,14]

The samples were characterized by time-of-flight (TOF) powder neutron diffraction using the Super-High-Resolution-Powder-Diffractometer (SuperHRPD)[15,16] at BL08 in J-PARC MLF. Note that ~2 g of samples used for neutron diffraction were placed in a V-Ni null cell made by Taiyokoko Co., LTD. having an inner diameter of 6 mm, and the diffraction profiles of the low-angle (LA), back-surface (BS), and 90° (QA) banks were obtained at room temperature. Crystallographic parameters were then refined using the Rietveld method using Z-Rietveld.

In this study, Al concentrations $x$ of samples for which the physical properties were measured were determined using the site occupancies of Al and Fe obtained from the Rietveld analysis of neutron and XRD data. In this manner, the concentrations determined are a few percent lesser than the nominal concentrations at which the samples were synthesized. This compositional shift is because a higher synthesis temperature caused the Al to sublimate.

In the applied magnetic field ($H$) between 0.01 and 2 T, the temperature ($T$) dependence of magnetization ($M$) was measured using a magnetic property measurement system (MPMS, Quantum Design) and a vibrating sample magnetometer (*MagLab*$^{VSM}$, Oxford Instruments). Magnetization measurements were conducted on heating after zero-field cooling (ZFC) and on subsequent cooling (FC). Isothermal magnetization was performed in the magnetic field range of ±7 T using MPMS. The temperature dependence of AC susceptibility was measured in oscillating magnetic fields of 0.39 mT at several frequencies between 10 and 1000 Hz using MPMS.

## 3. Results

*3.1 Characterization*

Here, we report the crystal properties of the sample with $x = 0.15$ as a representative of the synthesized samples. Figure 3 shows the XRD pattern of the sample with $x = 0.15$. All peaks shown in Fig. 3 can be attributed to the andalusite structure with the space group of *Pnnm*. Table I demonstrates the lattice parameters of the sample with $x = 0.15$ determined by the Rietveld refinement of XRD.

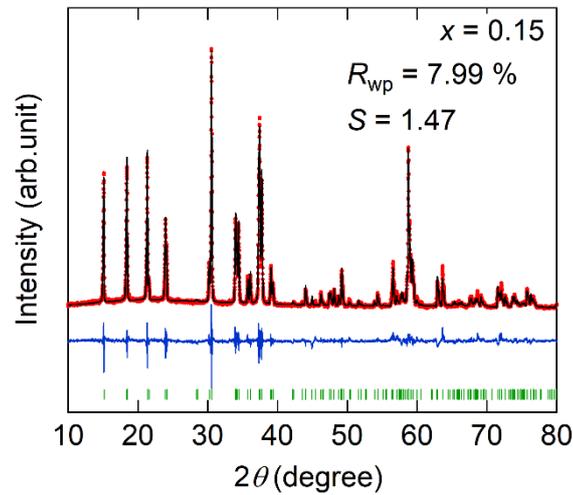

Fig. 3 (Color online) Powder XRD pattern of the sample with $x = 0.15$. The red points represent the observed intensity. The black and blue lines indicate the calculated and difference between observed and calculated intensities, respectively. The green vertical bars indicate the positions of Bragg reflections.

When a Fe-rich sample is measured using CuKα radiation, Fe atom generates fluorescent X-rays, which increases the background and makes the observed peak intensity relatively small.

Table I. Lattice parameters of the sample with $x = 0.15$.

| | |
|---|---|
| $a$ (Å) | 8.2174(1) |
| $b$ (Å) | 8.3115(1) |
| $c$ (Å) | 5.9043(1) |

Therefore, to determine the crystallographic parameters including the position of the oxygen atoms precisely, high-resolution neutron diffraction experiments were performed on samples with $0.09 \leq x \leq 0.91$. For example, Table II shows the lattice and crystallographic parameters of the sample with $x = 0.15$ determined by the Rietveld refinement of powder neutron diffraction. The obtained lattice parameters $a = 8.2166(3)$ Å, $b = 8.3110(3)$ Å, and $c = 5.9039(2)$ Å agree with the values obtained from XRD experiments. The Fe1 site has a larger $g$ value compared to the Fe2 site, indicating that $Fe^{3+}$ is more stable in the octahedral site compared to the trigonal bipyramidal site.

Table II. Crystallographic parameters of the sample with $x = 0.15$ determined by the Rietveld analysis of powder neutron diffraction data. $g$ and $B$ are the site occupancy and isotropic thermal displacement parameter, respectively. The orthorhombic lattice parameters are $a = 8.2166(3)$ Å, $b = 8.3110(3)$ Å, and $c = 5.9039(2)$ Å.

| Label | site | $g$ | $x$ | $y$ | $z$ | $B(Å^2)$ |
|---|---|---|---|---|---|---|
| Fe1 | 4$e$ | 0.973(1) | 0 | 0 | 0.24219(6) | 0.237(4) |
| Al1 | 4$e$ | 0.027(1) | 0 | 0 | 0.24219(6) | 0.237(4) |
| Fe2 | 4$g$ | 0.878(1) | 0.36615(4) | 0.13916(4) | 1/2 | 0.237(4) |
| Al2 | 4$g$ | 0.122(1) | 0.36615(4) | 0.13916(4) | 1/2 | 0.237(4) |
| Ge | 4$g$ | 1 | 0.24450(4) | 0.25079(4) | 0 | 0.131(5) |
| O1 | 4$g$ | 1 | 0.42622(5) | 0.35977(6) | 0 | 0.393(4) |
| O2 | 4$g$ | 1 | 0.09807(5) | 0.40152(5) | 0 | 0.393(4) |
| O3 | 4$g$ | 1 | 0.42403(5) | 0.35958(6) | 1/2 | 0.393(4) |
| O4 | 8$h$ | 1 | 0.22617(3) | 0.12718(4) | 0.24208(6) | 0.393(4) |

Figure 4(a) shows the $x$ dependence of Fe1 site occupancy $g_{Fe1}$ and Fe2 site occupancy $g_{Fe2}$ obtained from the Rietveld analysis of neutron diffraction data of $Al_xFe_{2-x}GeO_5$ ($0.09 \leq x \leq 0.91$). For any $x$, $g_{Fe1}$ is higher than $g_{Fe2}$, clearly indicating that $Fe^{3+}$ in the octahedral site is stable. This tendency agrees with the result of the previous ESR study.[17] Figure 4(b) shows the ratio of $g_{Fe2}$ to $g_{Fe1}$ as $x$ dependence. Indeed, this ratio decreases with increasing in $x$. For the samples with $x = 0.91$, the number of Fe in the trigonal bipyramidal sites is ~ 60% of Fe in octahedral sites. Moreover, this ratio does not appear to be one in the extrapolation to $x = 0$, i.e., at $x = 0$, there is less Fe in the trigonal bipyramidal sites compared to octahedral sites, although the number of octahedral and trigonal bipyramidal sites is crystallographically equal.

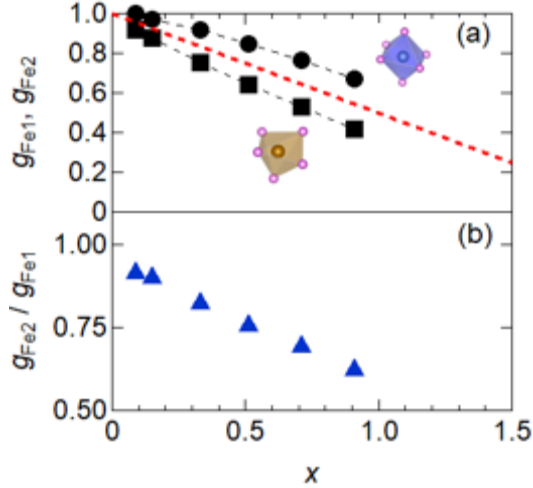

Fig. 4. (Color online) (a) Dependence of the occupancy of each Fe site on $x$ obtained from $Al_xFe_{2-x}GeO_5$ ($0.09 \leq x \leq 0.91$). Circles and squares indicate $g_{Fe1}$ (octahedral site) and $g_{Fe2}$ (trigonal bipyramidal site), respectively. The black dotted curves are a guide to the eye. The red dotted line shows the variation of the average value between $g_{Fe1}$ and $g_{Fe2}$ with respect to $x$. (b) The $x$ dependence of the ratio of $g_{Fe2}$ to $g_{Fe1}$.

For the andalusite structure, there are five types of superexchange interactions between the nearest magnetic ions. In Fig. 5, these five types are represented as bonds between magnetic ions. As shown in this figure, there are two types of Fe1–Fe1 bonds designated as $J_a$ and $J_b$, one type of Fe2-Fe2 bond designated as $J_c$, and two types of Fe1–Fe2 bonds designated as $J_d$ and $J_e$. Because all these are superexchange interactions between $Fe^{3+}$ ions through $O^{2-}$, they should be antiferromagnetic according to the Kanamori–Goodenough rule.

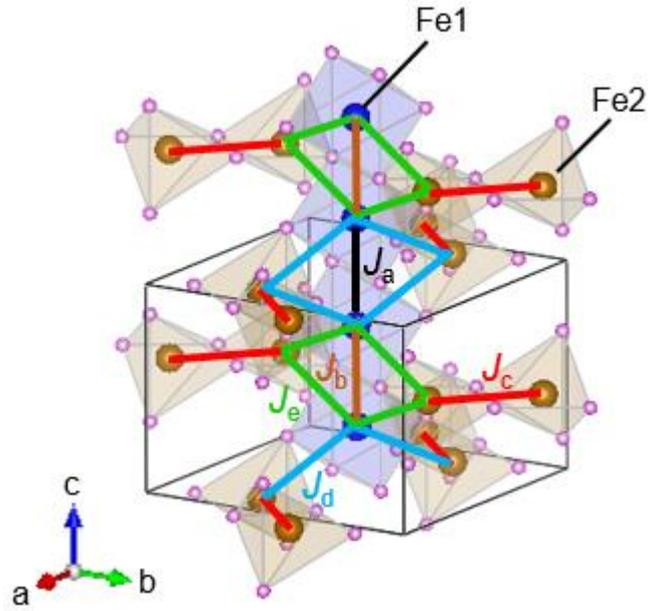

Fig. 5. (Color online) Five types of bonds between neighboring magnetic sites via oxygen in $Al_xFe_{2-x}GeO_5$: $J_a$ (black), $J_b$ (brown), $J_c$ (red), $J_d$ (blue), and $J_e$ (green). The pink spheres represent O. The blue spheres Fe1 are located in the octahedral sites and the brown spheres Fe2 in the trigonal bipyramidal sites. The solid black lines show the unit cell. The VESTA program is used for visualization.[4]

## 3.2 Magnetizations of $Al_xFe_{2-x}GeO_5$ with $0.09 \leq x \leq 0.26$

Figure 6(a) shows the temperature dependences of ZFC $M/H$ and FC $M/H$ for the samples with $x = 0.09$, 0.15, and 0.26 measured at $\mu_0H = 0.01$ T. $M/H$ for all samples exhibit similar temperature dependence. With a reduction in temperature, ZFC $M/H$ reaches its maximum at a certain temperature. We define this temperature as $T_{peak}$. Below a temperature slightly higher than $T_{peak}$, there are differences between ZFC $M/H$ and FC $M/H$; this difference increases as the temperature decreases and the FC $M/H$ remains constant at low temperatures. The behavior of the FC $M/H$–$T$ curve for each sample is similar to that of P-type ferrimagnets. $T_{peak}$ becomes lower as $x$ increases, reflecting the decrease in the number of magnetic Fe ions; $T_{peak} = 58$, 51, and 38 K for the samples with $x = 0.09$, 0.15, and 0.26, respectively; however, the $M/H$ value at $T_{peak}$ is highest for the sample with $x = 0.15$. This observation indicates that the magnetic response associated with the magnetic phase transition at $T_{peak}$ does not systematically change with $x$. Note that ZFC $M/H$ in the low-temperature region for the samples with $x = 0.15$ and 0.26 are slightly negative because of the presence of a residual magnetic field of $\mu_0H \sim -0.001$ T during cooling. As shown in Fig. 6(b), the inversed susceptibility FC $H/M$ for the samples with $x = 0.09$, 0.15, and 0.26 measured at $\mu_0H = 0.1$ T show convex temperature dependences at high temperatures, indicating the short-range spin correlations exist up to a temperature at 300 K. If the temperature dependence of FC $M/H$ is measured up to higher temperatures and analyzed using the Curie–Weiss (CW) law, a negative value of Weiss temperature $\theta_W$ can be determined. The negative values of $\theta_W$ indicate that the dominant magnetic interaction is antiferromagnetic, which agrees with the Kanamori–Goodenough rule.

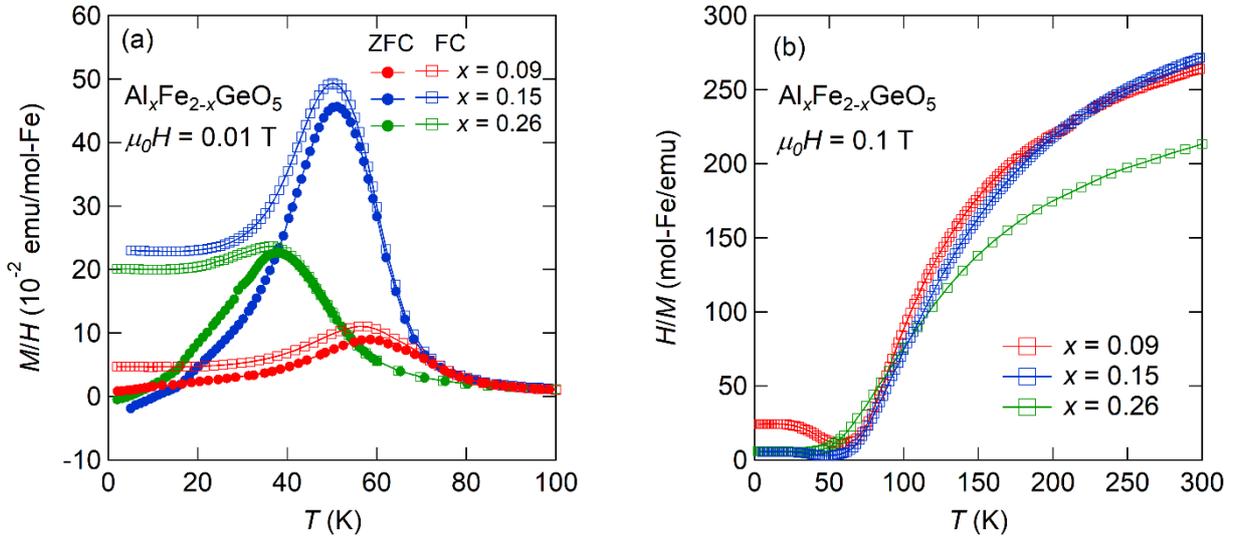

Fig. 6. (Color online) Temperature dependences of (a) $M/H$ measured under ZFC and FC conditions, and (b) FC $H/M$ for samples with $x = 0.09$, 0.15, and 0.26.

Figure 7 shows the isothermal magnetization at 4.2 K for the sample with $x = 0.15$ as a representative of these samples. A small amount of remanent magnetization is observed, indicating the presence of a small spontaneous magnetic moment. Because of the negative value of $\theta_W$ and small spontaneous magnetization, the sample is possibly a P-type ferrimagnet. From the results of magnetization measurements, it is impossible to determine the magnetic state of these samples at low temperatures. However, due to the presence of a small amount of spontaneous magnetization, we call this state the weak ferromagnetic-like state for convenience.

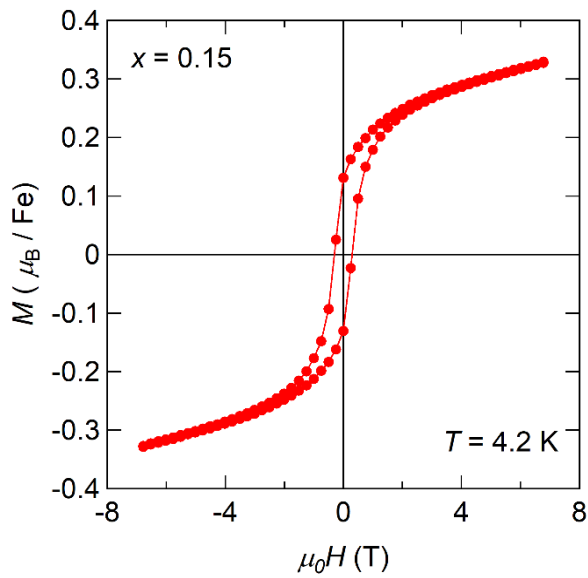

Fig. 7. (Color online) Isothermal magnetization $M$ at 4.2 K for the sample with $x = 0.15$.

Figure 8 shows the temperature dependences of ZFC $M/H$ and FC $M/H$ for the sample with $x = 0.26$ measured in several magnetic fields. With increasing in the magnetic field, the peak of ZFC $M/H$ shifts to lower temperatures with broadening and disappears at $\mu_0 H = 0.5$ T. This behavior seems to be similar to that observed in spin-glasses; however, the measurement of the AC susceptibility $\chi'$ clarifies that this sample is not a spin-glass. Figure 9 shows the temperature dependences of AC $\chi'$ for the sample with $x = 0.26$ measured in several frequencies. No frequency dependence was observed in the temperature variation of $\chi'$. For all frequencies, $\chi'$ shows a peak at almost the same temperature as $T_{peak}$ determined by ZFC $M/H$ measurement. From the $\chi'$ measurement results, we concluded that the difference between the temperature dependence of $M/H$ observed in the ZFC condition and that in the FC condition is not caused by spin fluctuations as in spin-glass but is caused by the hysteresis loop of the $M$–$H$ curve as shown in Fig. 7.

From the results of above magnetization measurements, we considered that $Al_xFe_{2-x}GeO_5$ with $x$ = 0.09, 0.15, and 0.26 are weak ferromagnetic-like samples.

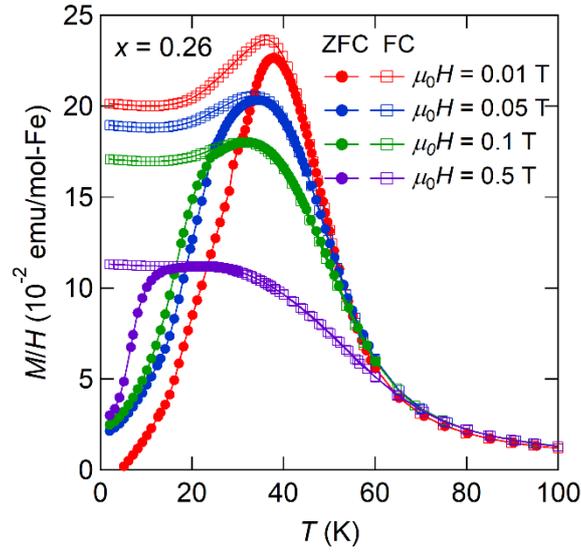

Fig. 8. (Color online) Temperature dependences of $M/H$ for the sample with $x$ = 0.26 measured under several magnetic fields.

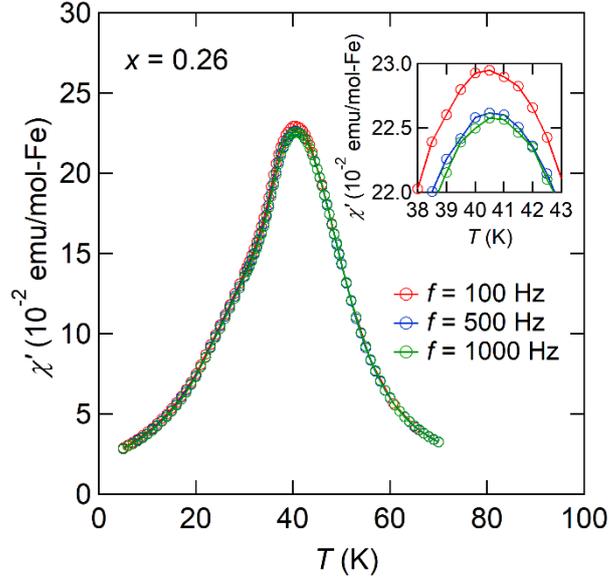

Fig. 9. (Color online) Temperature dependences of AC $\chi'$ for the sample with $x = 0.26$ at several frequencies. The inset shows expanded $\chi' - T$ curve around 40 K.

*3. 3 Magnetizations of $Al_xFe_{2-x}GeO_5$ with $0.33 \leq x \leq 0.91$*

Figure 10(a) shows the temperature dependence of ZFC $M/H$ and FC $M/H$ for the samples with $x = 0.33$, 0.51, 0.71, and 0.91 measured in $\mu_0 H = 0.05$ T. ZFC $M/H$ has a maximum value at temperature $T_{peak}$. Below $T_{peak}$, ZFC $M/H$ becomes smaller than FC $M/H$; as the temperature reduces, the difference between ZFC and FC $M/H$ increases. These are similar to the behaviors observed in spin-glasses. The temperature dependences of FC $H/M$ for the samples with $x$ = 0.33, 0.51, 0.71, and 0.91 measured at $\mu_0 H$ = 0.1 T are shown in the Fig. 10(b). The $H/M$ of all samples show an upward convex temperature dependence above $T_{peak}$, indicating the presence of short-range spin correlations up to 300 K and above.

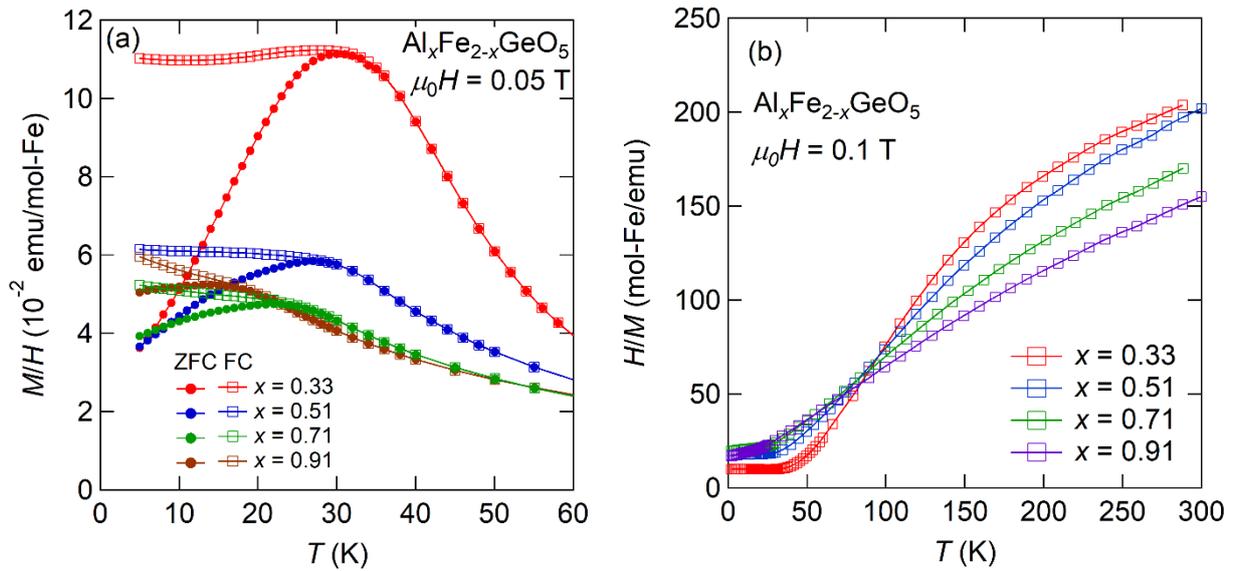

Fig. 10. (Color online) Temperature dependences of (a) $M/H$ measured under ZFC and FC conditions, and (b) FC $H/M$ for the samples with $x = 0.33$, 0.51, 0.71, and 0.91.

Figure 11 shows the isothermal magnetization at 2 K for the sample with $x$ = 0.51 as a representative of these samples. The isothermal magnetization gradually increases with an increasing magnetic field. Moreover, there is a narrow hysteresis loop below approximately 4

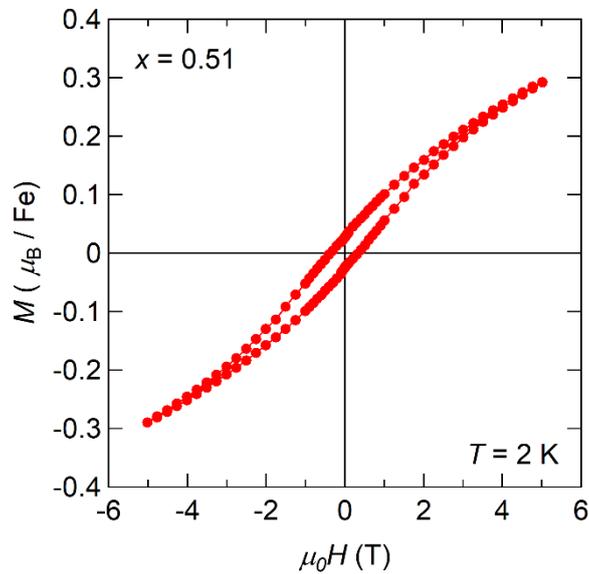

Fig. 11. (Color online) Isothermal magnetization $M$ at 2 K for the sample with $x$ = 0.51.

T. This field dependence of the isothermal magnetization, which appears to be absent of spontaneous magnetization, is exceedingly different from that of the sample with $x$ = 0.15 shown in Fig. 7.

Figure 12 shows the temperature dependences of ZFC $M/H$ and FC $M/H$ for the samples with $x = 0.51$ and $x = 0.91$ measured at multiple magnetic fields. As the applied field increases, the peak of ZFC $M/H$ appearing at $T_{peak}$ shifts to lower temperatures with broadening. For the sample with $x = 0.51$, it becomes plateau-shaped at $\mu_0 H = 2$ T. On the other hand, $T_{peak}$ exists even at $\mu_0 H = 2$ T for the sample with $x = 0.91$. In other words, the response of the magnetization in the sample with $x = 0.91$ to the magnetic field is weaker than that in the sample with $x = 0.51$.

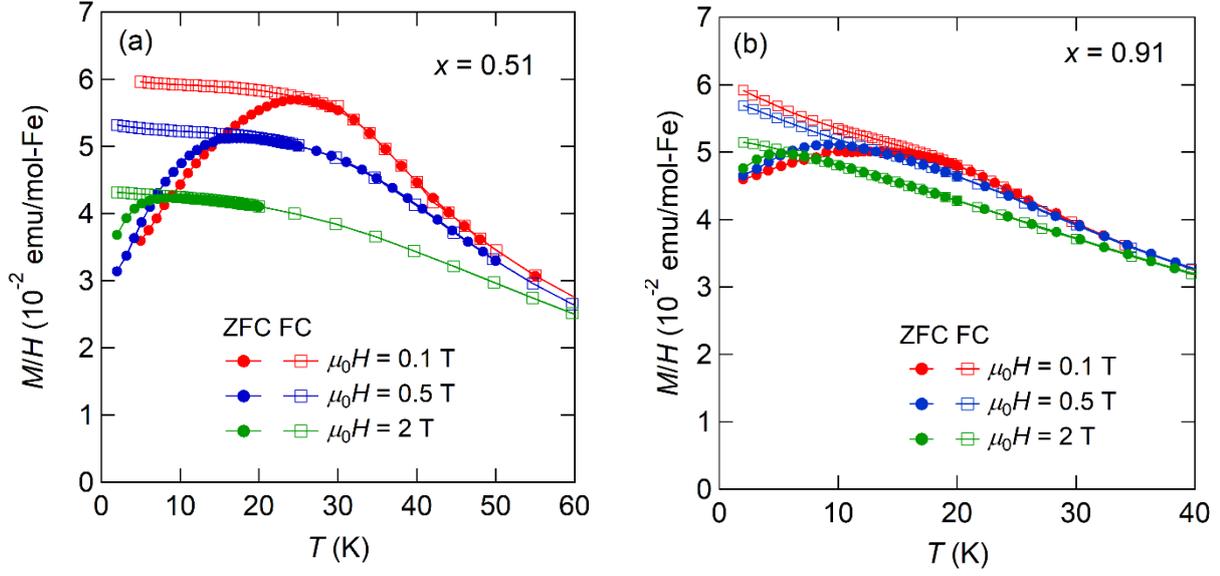

Fig. 12. (Color online) Temperature dependences of $M/H$ for the samples with (a) $x = 0.51$ and (b) $x = 0.91$ under several magnetic fields.

Figure 13 shows the temperature dependences of AC $\chi'$ for the sample with $x = 0.51$ and $x = 0.71$ measured at several frequencies. At $f = 10$ Hz, the peak of $\chi'$ for each sample appears at a temperature $T_f(f)$ slightly higher than $T_{peak}$ determined from the temperature dependence of ZFC $M/H$. The reason for differences in the peak temperatures is that $T_{peak}$ was determined under a magnetic field of $\mu_0 H = 0.05$ T. The $T_f(f)$ for each sample shifts to a higher temperature side with increase in frequency similar to spin-glass. The parameters $\Delta T_f(f)/[T_f(f)\Delta \log f]$ are found to be 0.008 and 0.007 for the sample with $x = 0.51$ and $x = 0.71$, respectively. These are typical values for metallic spin-glasses but are considerably smaller than those for superparamagnets, e.g., 0.28 for a-$(Ho_2O_3)(B_2O_3)$.[18]

Based on the abovementioned magnetization measurements, $Al_xFe_{2-x}GeO_5$ with $x = 0.33$, 0.51, 0.71, and 0.91 are spin-glasses.

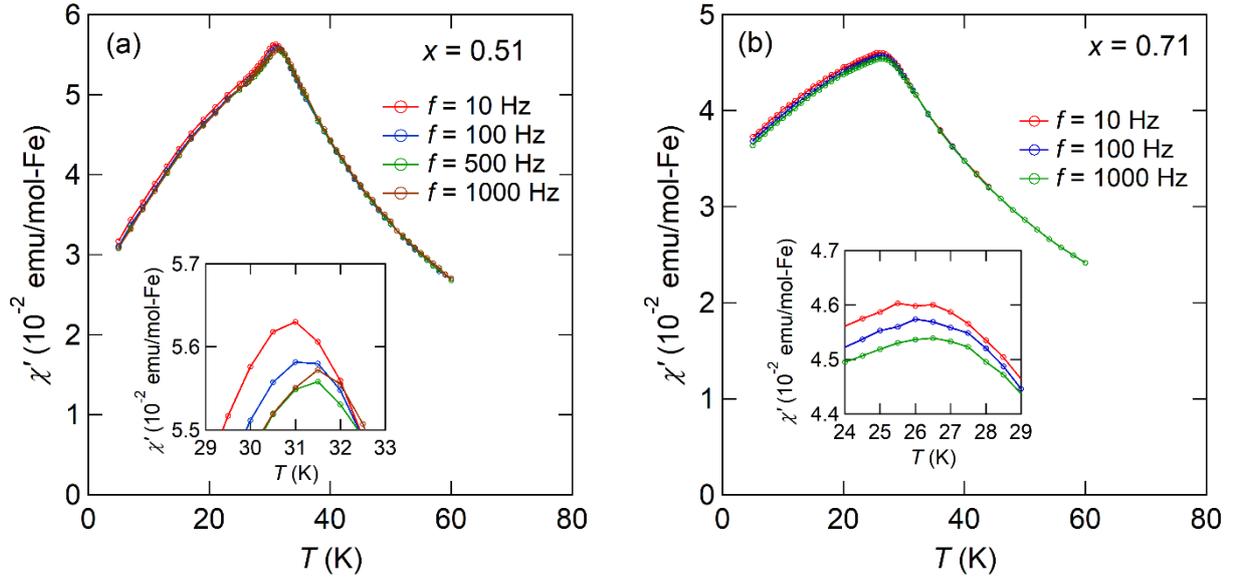

Fig. 13. (Color online) Temperature dependences of AC $\chi'$ for the samples with (a) $x = 0.51$ and (b) $x = 0.71$ at several frequencies. The inset in each figure shows expanded $\chi' - T$ curve around the peak of $\chi'$.

## 4. Discussion

The sample with $x = 0.15$ shows small spontaneous magnetization at 4.2 K. As shown in Fig. 5, the Dzyaloshinsky–Moriya (DM) interaction between $Fe^{3+}$ spins occurs because of the broken spatial inversion symmetry between Fe1 and Fe2 sites. In general, $Fe^{3+}$ in the octahedral sites has $S = 5/2$. In contrast, $Fe^{3+}$ in the trigonal pyramidal sites is expected to be in an intermediate state with $S = 3/2$.[19ced] Therefore, the origin of the presence of small spontaneous magnetization at low temperatures is unknown at this stage. However, it may be due to the existence of DM interaction or the difference in spin magnitude between Fe1 and Fe2 sites. In the future, the magnetic states of these samples at low temperatures will be clarified by neutron scattering experiments.

As shown in Fig. 10(a), the samples with $x = 0.33, 0.51, 0.71$, and $0.91$ are considered to be spin-glass magnets. This result agrees with previous studies.[11,12] Usually, the spin-glass state is attributed to the competition between ferromagnetic and antiferromagnetic interactions. In $Al_xFe_{2-x}GeO_5$, spin-glasses are considered to be generated by the same mechanism as in the insulating diluted magnetic materials. That is, dilution may cause competition between the nearest-neighbor interaction and the next-nearest-neighbor interaction.

From the temperature dependence of ZFC $M/H$, we define $T_{peak}$ observed for the samples with $x \leq 0.26$ as the weak ferromagnetic-like transition temperature $T_C$; $T_C$ = 58, 51, and 38 K for samples with $x$ = 0.09, 0.15, and 0.26, respectively. Also, we define $T_{peak}$ observed for the samples with $x \geq 0.33$ as the spin-glass transition temperature $T_g$; $T_g$ = 30 K, 27 K, 22 K, and

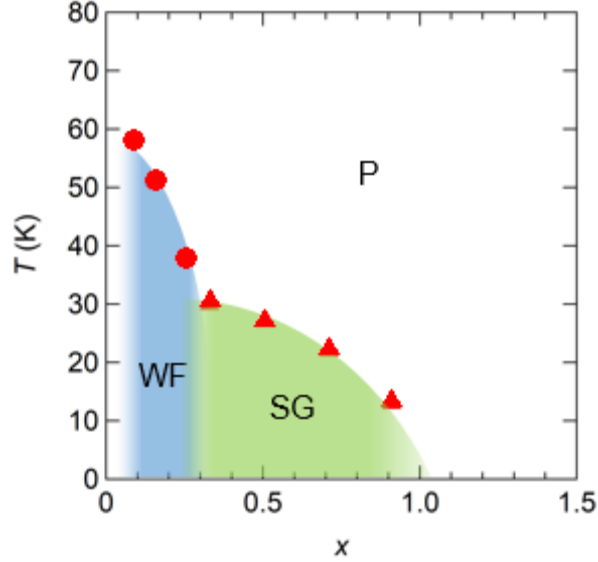

Fig. 14. (Color online) The $x$ versus $T$ phase diagram of andalusite-type $Al_xFe_{2-x}GeO_5$. WF: weak ferromagnetic-like phase, SG: spin-glass phase, P: paramagnetic phase.

13 K for samples with $x$ = 0.33, 0.51, 0.71, and 0.91, respectively. Fig. 14 shows the $x$ dependence of magnetic transition temperatures $T_C$ and $T_g$. There appears to be a clear phase boundary between WF and SG phases at around $x$ = 0.3.

In general, for random diluted magnetic systems, percolation is strongly related to establishing the magnetic order. In such systems, long-range order occurs when the concentration of magnetic elements is greater than the percolation threshold $p_c$. The $p_c$ values have been calculated for site-random and bond-random cases for multiple crystal lattices, and comparisons with the percolation threshold occurring in real systems have been discussed.[20] From numerical calculations, the site-percolation threshold value of the three-dimensional stacked bow-tie lattice is reported to be 0.2822.[21] The $p_c$ of andalusite structure has not been calculated; however, using $x$ = 0.3, where the phase boundary between WF and SG is located in the $x$–$T$ phase diagram, the $p_c$ of $Al_xFe_{2-x}GeO_5$ is estimated to be $p_c$ = $(2-x)/2$ = 0.85. This increase in percolation shows that although the crystal structure of $Al_xFe_{2-x}GeO_5$ is somewhat similar to a stacked bow-tie lattice, as shown in Fig.2, a little dilution can block the path of magnetic interaction. If magnetic interactions indicated by $J_a$ and $J_b$ are dominant among the five types of magnetic interactions shown in Fig. 5, i.e., if the magnetic order is formed in $Al_xFe_{2-x}GeO_5$, which can be considered as a one-dimensional magnetic chain, the magnetic order is prevented by a slight dilution. In the future, this andalusite structure will be of interest for percolation problems.

As shown in Fig. 4(b), the ratio of Fe occupancy of trigonal bipyramidal sites to that of octahedral sites gradually decreases as $x$ increases. This indicates that $J_c$ tends to disappear with ease of increasing in $x$ than $J_a$ and $J_b$ in Fig. 5 and that the mechanism of competition between magnetic interactions may change with $x$. This is probably the reason for the different magnetic field dependence of the temperature change of $M/H$ between the samples with $x = 0.51$ and $x = 0.91$. If magnetic interactions in $Al_xFe_{2-x}GeO_5$ are clarified by neutron scattering experiments, the mechanism of weak ferromagnetism, the competition mechanism of magnetic interactions in spin-glasses, and the limitation of percolation threshold will be revealed.

Finally, we discuss the possibility of polymorphic magnets $Fe_2GeO_5$ ($x = 0$). Figure 4(b) shows that even at $x = 0$, the number of Fe ions located in octahedral and trigonal bipyramidal sites differ, i.e., $Fe_2GeO_5$ with andalusite structure is not formed. On the other hand, in the kyanite structure, as shown in Fig. 1(b), all the Fe ions are located in the octahedral sites, which indicates that the synthesis of $Fe_2GeO_5$ with kyanite structure is possible. At present, neither structure of $Fe_2GeO_5$ has been successfully synthesized via the solid-state reaction method. To improve our understanding of polymorphic magnets, the synthesis of $Fe_2GeO_5$ is a challenge for the future.

## 5. Conclusions

Polycrystalline $Al_xFe_{2-x}GeO_5$ ($0.09 \leq x \leq 0.91$) was synthesized via a solid-state reaction method under vacuum-like conditions. The substitution rate of Fe ions in the octahedral and trigonal bipyramid sites, however, were different. The magnetism of $Al_xFe_{2-x}GeO_5$ at low temperatures changes with increasing in $x$ from a weak ferromagnetic-like state to a spin-glass state; the phase boundary between two states seems to be located around $x = 0.3$. The origin of the presence of small spontaneous magnetization at low temperatures in the samples with $x < 0.3$ may be the existence of DM interaction or the difference in spin magnitude between $Fe^{3+}$ in the octahedral and trigonal pyramidal sites. The appearance of spin-glass in the samples with $x > 0.3$ suggests that the dilution of Fe ions by Al ions leads to competition between ferromagnetic and antiferromagnetic interactions. If $x = 0.3$ is considered as the percolation limit, the magnetic materials with andalusite structure will lose long-range order due to substitution by fewer nonmagnetic ions than those with stacked bow-tie lattice. Based on the $x$ dependence of the site occupancy of Fe ions, it is expected that $Fe_2GeO_5$ with andalusite structure is not formed.

**Acknowledgment**

We thank Masaki Kato for his help with the magnetization measurements. This work was supported by the Japan Society for the Promotion of Science (JSPS) KAKENHI Grants Nos. JP19K14646, JP18K03506, and JP21K03441. A part of this study was carried out by the joint research in the Institute for Solid State Physics, the University of Tokyo. The neutron experiment at the Materials and Life Science Experimental Facility of the J-PARC was performed under a user program (Proposal No.2014S05).

*s194819y@st.go.tuat.ac.jp

†h-katori@cc.tuat.ac.jp

‡Present address: Multiple-Degree-of-Freedom Correlation Research Group, Neutron Materials Research Division, Materials Science Research Center, Japan Atomic Energy Agency, Tokai, Ibaraki 319-1195, Japan

✶Present address: Spallation Neutron Source Science Center, Dongguan, 523803, China